

\PHYSREV

\titlepage
\nopubblock
\title{Measurement of the Schr\"odinger wave of a single particle}
\author{Yakir Aharonov and Lev Vaidman}

\address{
School of
Physics and Astronomy \break
Raymond and Beverly Sackler Faculty of Exact Sciences \break
Tel-Aviv University, Tel-Aviv, 69978 ISRAEL}

\andaddress{ Physics Department, University of South Carolina \break
Columbia, South Carolina 29208, U.S.A.}
\abstract

We show that it is possible to measure the Schr\"odinger wave of a
single quantum system.  This provides a strong argument for
associating physical reality with the quantum state of a single
system, and challenges the usual assumption that the quantum state has
physical meaning only for an ensemble of identical systems.

\endpage

While the Schr\"odinger wave is a basic element of quantum theory, it
is generally believed that one cannot associate physical reality to
the wave of a single particle.  Rather, the Schr\"odinger wave is
often viewed as a mathematical tool for calculating the probabilities
of various outcomes of certain experiments when many such experiments
are performed on an ensemble of identical systems, all in the same
quantum state.  Several arguments seem to support this point of view:

i) We have never seen the quantum state of a single particle in a
laboratory.  Indeed, while a wave is typically spread over a region of
space, we never see a particle simultaneously in several distinct
locations.

ii) If we could see a quantum state, we could presumably distinguish
it from any other quantum state, but the unitary time evolution of
states in quantum mechanics implies that it is impossible to
distinguish between two different non-orthogonal states.  Different
outcomes of a measurement to distinguish these two states correspond
to orthogonal quantum states of the composite system (measuring
device plus particle).  But, the initial scalar product between the
states was not zero and remains nonzero under unitarity time
evolution.

iii) If we associate  physical reality with a spread-out wave then
the instantaneous ``collapse" of the wave to a point during a
position measurement seems to conflict with relativity [1].

In this letter we demonstrate how the density $\rho = \Psi^* \Psi$ and
the current $j = {{\hbar}\over{2mi}}( \Psi^* \nabla \Psi - \Psi \nabla
\Psi^*)$ of a Schr\"odinger wave of a single particle can be measured.
The usual measurements assumed in argument (i) alter the Schr\"odinger
wave and are not adequate; rather, here we will describe special {\it
protective} measurements allowing us to measure $\rho$ and $j$ without
changing the Schr\"odinger wave.  In some cases energy conservation
provides protection for the state, while in other cases we need a
special protection procedure.

Let us consider a particle in a discrete nondegenerate energy
eigenstate $\Psi (x)$. The standard von Neumann
procedure for measuring the value of an observable $A$ in
this state involves an
 interaction Hamiltonian
$$
 H = g(t) p A,\eqno(1)
$$
coupling our system to a measuring device, or pointer, with coordinate
and momentum denoted respectively by $q$ and $p$.  The time-dependent
coupling $g(t)$ is normalized to $\int g(t) dt =1$.  The initial state
of the pointer is taken to be a Gaussian centered around zero.

In standard impulsive measurements, $g(t) \neq 0 $ only for a very
short time interval.  Thus, the interaction term dominates the rest of
the Hamiltonian, and the time evolution
$
e^{-{i\over \hbar}p A}
$
leads to a correlated state: eigenstates of $A$ with eigenvalues $a_n$
are correlated to measuring device states in which the pointer is
shifted by these values $a_n$.  By contrast, the protective
measurements of interest here utilize the opposite limit of extremely
slow measurement.  We take $g(t) = 1/T$ for most of the time $T$ and
assume that $g(t)$ goes to zero gradually before and after the period
$T$. We choose the initial state of the measuring device  such
that the canonical conjugate $p$ (of the pointer variable $q$) is
bounded.  For $g(t)$ smooth enough we obtain an adiabatic process in
which the particle cannot make transition from one energy eigenstate
to another, and, in the limit $T \rightarrow \infty$, the interaction
Hamiltonian does not change the energy eigenstate.  For any value of
$p$, the energy of the eigenstate shifts by an infinitesimal amount
given by first order perturbation theory:
$$\delta E = \VEV{H_{int}} = {{\VEV{A}p}\over T}. \eqno(2)$$
The corresponding time evolution $ e^{-i p \VEV{A}} $ shifts the
pointer by the average value $\VEV{A}$.  (Here and below we will take
$\hbar =1$.)  This result contrasts with the usual (strong)
measurement in which the pointer shifts by one of the eigenvalues of
$A$.  By measuring the averages of a sufficiently large number of
variables $A_n$, the full Schr\"odinger wave $\Psi (x)$ can be
reconstructed to any desired precision.

As a specific example we take the $A_n$ to be (normalized) projection
operators on small regions $V_n$  having volume $v_n$:

$$A_n=\cases{ {1\over {v_n}},&if $x \in V_n$;\cr
0,&if $x \not\in V_n$.\cr} \eqno(3)
$$

\noindent
 The measurement of $A_n$ yields $$\VEV{A_n} = {1\over {v_n}}
\int_{V_n} |\Psi|^2 dv = |\Psi_n|^2 ,\eqno(4) $$ where $ |\Psi_n|^2 $
is the average of the density $\rho(x) = |\Psi(x)|^2$ over the small
region $V_n$.  Performing measurements in sufficiently many regions
$V_n$ we can reconstruct $\rho(x)$ everywhere in space.  (Simultaneous
measurement of all the variables $A_n$ requires slower and weaker
interactions, and thus takes more time.)  For a real state the density
$\rho(x)$ is itself enough to reconstruct the Schr\"odinger wave; we
can fix the sign by flipping it across nodal surfaces.

In general case, however, in addition
to measurements of the density $\rho(x)$, we have to measure current
density.  This time we also adiabatically measure the averages of
$$
B_n ={1\over{2i}} (A_n\nabla + \nabla A_n)  ~~~.\eqno(5)
$$
Indeed, $\VEV{B_n}$ are the average values of the current
$j = {1\over{2im}} (\Psi^* \nabla \Psi - \Psi  \nabla \Psi^* )$ in the
region $V_n$. Writing $\Psi (x) = r(x) e^{i\theta(x)}$ with
$r(x)=\sqrt{\rho (x)}$,
 we find that
$$
{{m j(x)} \over {\rho (x)}} = \nabla \theta ~~~;\eqno(6)
$$
and  the phase $\theta (x)$ can be found by integrating
$j/\rho$.

For a charged particle the density $\rho (x)$ times the charge yields
the effective charge density. In particular, it means that an
appropriate adiabatic measurement of the Gauss
flux out of a certain region must yield the expectation value of the
charge inside this region (the integral of the charge density over
this region). Likewise, adiabatic measurement of the Ampere contour integral
yields the expectation value of the total current flowing through this
contour in the stationary case.

Our discussion of the current of the particle is valid only for a
Hamiltonian without vector potential.  However, the eigenstates of
such a Hamiltonian with a nonvanishing current are necessarily
degenerate due to time reversal invariance.  The method described
above is appropriate only for nondegenerate eigenstates and,
therefore, we have to consider problems with a vector potential
$\bf{A}$, for which we do have nondegenerate stationary states with
non-zero current (e.g. the Aharonov-Bohm effect).  Then, the
definition of the (electric) current must be modified by replacement
$\nabla \rightarrow \nabla - ie{\bf A}$.  This replacement has to be
done also for the definition of the observables $B_n$ (Eq.(5)), and it
leads to the obvious modification of the Eq.(6).

We have shown that stationary quantum states can be observed.  This
is our main argument for associating physical reality with the quantum
state of a single particle.  Since our measurement lasts a long period
of time we do not have a method for measuring the Schr\"odinger wave
at a given time.  Thus, we have a direct argument for associating
physical reality with stationary Schr\"odinger waves only over a {\it
period} of time.  The reader may therefore suspect that our
measurements represent time-averaged physical properties of the
system.  Let us now present a few arguments explaining why,
nevertheless, these
measurements reflect properties of the Schr\"odinger wave at any given
moment of time during the measurement.

An essential feature of our adiabatic measurement is that the state
$|\Psi \rangle$ does not change throughout the experiment.  Since the
Schr\"odinger wave yields the complete description of a system and the
interaction with the measuring device is constant throughout the
measurement, we conclude that the action of the system on the
measuring device is the same at any moment during the measurement.

The mathematical description of our measurement tells us the same: for
any, even very short, period of time, the measuring device shifts by
an amount proportional to $\langle A \rangle $, the expectation value
of the measured variable, rather than to one of its eigenvalues $a_n$.
Thus, expectation values, which mathematically characterize
Schr\"odinger waves, can be associated with very short periods of
time.  In the instantaneous limit, expectation
values and,
therefore, the quantum state manifest themselves as properties of a
quantum system defined at a given time.  (Note, however, that
pointer shifts during short time intervals are unobservable since
they are much smaller than the uncertainty; only the total shift
accumulated during the whole period of measurement is much larger than
the width of the initial distribution, and therefore observable on a
single particle.)

Moreover, suppose that (contrary to standard quantum theory), a system
has a complete description that {\it does} change during the
measurement process, and the (constant) Schr\"odinger wave we measure
does not describe the system at a given time but represents only a
time average of some hidden variables over the period of the
measurement.  Consider a model of a hydrogen atom in which the
electron performs very fast ergodic motion in the region corresponding
to the quantum cloud.  The charge density might be either zero (if the
electron is not there) or singular (if the electron is inside the
infinitesimally small region including the space point in question).
In spite of this fact, the measurement we have described will yield
outcomes corresponding to a nonsingular charge density cloud.  What it
measures is the time average of the density, or how long a time the
electron spent in a given place.

In order to see that this picture is inappropriate for the quantum
case let us consider another example: a particle in a one-dimensional
box of length $L$ in the first excited state.  The spatial part of the
state is $\sqrt{2/L} \sin (2\pi x/L)$.  The adiabatic measuring
procedure described above will yield the Schr\"odinger wave density
$(2/L) \sin^2 (2\pi x/L)$.  In particular, it equals zero at the
center of the box.  If there is some hidden position of the electron
which changes in time such that the measured density is proportional
to the amount of time the electron spends there, then half of the time
it must be in the left half of the box and half of the time in the
right half of the box.  But it can spend no time at the center of the
box; i.e., it must move at infinite velocity at the center.  It is
absolutely unclear what such an electron ``position" would be.  There
{\it is} a theory [2] which introduces a ``position" for a particle in
addition to its Schr\"odinger wave; but according to this theory, the
``velocity" of the particle in the given energy eigenstate vanishes:
it does not move at all.  In the quantum picture the eigenstate of the
particle in the box can be represented as a superposition of two
running waves moving in opposite directions.  The zero density at the
center of the box is due to destructive interference -- the phenomenon
which cannot be reproduced in a classical ergodic model of a particle.

The procedure described above cannot measure properties of a state
obtained by superposing several nondegenerate energy eigenstates.
Applied to such a state, a measurement of $A$ will yield shifts of the
pointer corresponding to the expectation values of the variable $A$ in
the various energy eigenstates.  In general, these values are distinct
with differences greater than the initial uncertainty of the pointer
position.  Thus, after the interaction, the system and the measuring
device are entangled.  By ``looking" at the measuring device we cause
the Schr\"odinger wave to choose one of the energy eigenstates.
Measurement of the Schr\"odinger wave -- namely, measurement of the
expectation values of the projection operators -- causes collapse.  A
superposition of nondegenerate energy eigenstates is not protected by
energy conservation: unitary evolution during the measurement leads to
correlations between energy states and the states of the measuring
device without changing the total energy, while collapse changes the
energy itself.

Nevertheless, we can measure even a superposition of energy
eigenstates by a procedure similar to the one described above.  We
just have to add an appropriate protection mechanism.  The simplest
way to protect a time-dependent Schr\"odinger wave is via dense
state-verification measurements that test (and thus protect) the time
evolution of the quantum state.  If we are interested in all the
details of this time-dependent state we cannot use measurements which
are too slow.  Every measurement of the density and current of a
Schr\"odinger wave must last a period of time which is smaller then
the characteristic time of the evolution of the state; and the time
intervals between consecutive protections must be even
smaller.  However, in principle, Schr\"odinger wave measurement to any
desired accuracy is possible: for
any
desired accuracy
 there is a density of the
state-verification
measurements that will protect the state from being changed due to the
measurement interaction.  Additional protection is necessary also for
measurement of stationary but degenerate states; and the scheme of
dense projection measurements is applicable here too.
Even for dense projective measurements, most of the time the system
evolves according to its free Hamiltonian, so we are allowed to say
that what we measure is the property of the system and not of the
protection procedure.

When measurements involve the above kind of protection, we have to
know the
state in order to prescribe the proper protection.  One might object,
therefore, that our measurement yields no new information, since the
state is already known.  However, we can separate the protection and
measurement procedures: one experimentalist provides protection and
the other measures the Schr\"odinger wave itself.  Then the second
experimentalist does obtain new information.  The most important
point, however,
is that we directly measure properties of the Schr\"odinger wave of a
single system using a standard measuring procedure.  Our direct
measurements of the density and the current of the Schr\"odinger wave
challenge the commonly accepted notion that quantum states can be
observed fully only when the measurement is performed on an ensemble of
identical systems.

Consider now an apparent paradox arising from the measurement of
Schr\"o-dinger wave.  It is well known that even assuming instantaneous
``collapse" of a quantum state, one cannot use the collapse for
sending signals faster than light.  At first, however, the possibility
of measuring the value of the Schr\"odinger wave at a given location
seems to allow such superluminal communication.  Consider a particle
in a superposition ${1\over\sqrt{2}} (|1\rangle + |2\rangle )$ of
being in two boxes separated by a very large distance.  For this
particle the expectation value of the projection onto the first box is
$\langle P_1 \rangle =1/2$.  This value must be the outcome of a
measurement performed on the first box.  If, however, just prior to a
measurement of the Schr\"odinger wave in the first box, someone opens
and looks into the second box, causing collapse to a localized state
$|1\rangle$ or $|2\rangle$, then the outcome of the measurement of the
projection operator in the first box will drastically change: we no
longer find $\langle P_1 \rangle = 1/2$ but rather 0 or 1 (depending
on what is found in the second box).  It seems, therefore, that
measurements on one box can influence measurements on another box
located arbitrarily far away.

However, this argument contains a flaw: the state ${1\over\sqrt {2}}
(|1\rangle + 2\rangle )$ is not a discrete nondegenerate eigenstate.
Since there is no overlap between the states $|1\rangle$ and
$|2\rangle$, the orthogonal state ${1\over\sqrt{2}} (|1\rangle -
|2\rangle)$ has the same energy.  Thus, there is no natural protection
due to the energy conservation, and an additional protection is
needed. This protection, however, involves
explicitly nonlocal interactions.  These nonlocal interactions are the
source of the alleged superluminal signal propagation.  (A more subtle
paradox of this sort is considered in another work [3].)

Let us come back to the three arguments against the realistic view of
the Schr\"odinger wave presented in the beginning of the letter.
First, we have shown that we {\it can} observe a quantum state.
Although our discussion relied on Gedanken experiments, recent
experimental work with so-called ``weak links" in quantum circuits
shows that slow adiabatic measurements of the Schr\"odinger wave can
be performed in the laboratory [4].

The second argument is a correct statement, but it only implies that
there is no single {\it universal} procedure for observing states.  It
is still allows for the possibility of an appropriate measuring
procedure for any given state.

The last argument (iii) is the most serious one.  Assume that the
Schr\"odinger wave of particle is nonvanishing only inside two
separate boxes, and we find it in one of them.  How did part of the
wave move instantaneously from one box to another?  We believe that a
full answer to this argument requires a new approach to quantum theory
[5].

We have shown that expectation values of quantum variables and the
quantum state itself have physical meaning, i.e., they are measurable
for individual quantum systems.  This result stands in sharp contrast
to the standard approach in which the Schr\"odinger wave and
expectation values are statistical properties of ensembles of
identical systems.

\vskip 0.8cm
\noindent
{\bf Acknowledgements}
\vskip 0.32cm

It is a pleasure to thank Sidney Coleman, Shmuel Nussinov, Sandu
Popescu and Daniel Rohrlich for helpful discussions.
 The research was supported in part by  grant 425/91-1 of the the
Basic
Research Foundation (administered by the Israel Academy of Sciences and
Humanities) and by grant PHY 8807812 of the National Science
Foundation.
\vskip 0.8cm
\noindent
\vskip 1cm
{\bf  References}
\vskip 0.32cm

\noindent
1~~Y. Aharonov and D. Albert, {\it Phys. Rev.} {\bf D24} (1981)
359.\hfill \break
2~~D. Bohm, {\it Phys. Rev.} {\bf 85} (1952) 166. \hfill  \break
3~~Y. Aharonov and L. Vaidman, {\it Proceedings of} ISQM-SAT, H. Ezawa
and Y. Murayama eds., {\it North-Holland Press}
 Tokyo (1992). \hfill  \break
4~~T.P. Spiller, T.D. Clark, R.J. Prance, and A.
Widom, {\it Prog. Low Temp. Phys.}  XIII \phantom{5~~}(1992) 219.
\hfill \break
5~~Y. Aharonov and L. Vaidman, {\it Phys. Rev.} {\bf
A41} (1990) 11.

\end